\documentclass[prd,twocolumn,amsmath,amssymb,floatfix,superscriptaddress]{revtex4-2}

\usepackage{times,graphics}
\usepackage[colorlinks]{hyperref}
\usepackage{makecell}
\usepackage{diagbox}

\usepackage{graphicx}
\usepackage{dcolumn}
\usepackage{bm}
\usepackage{natbib}
\usepackage{graphicx}
\usepackage{epsfig}
\usepackage{epstopdf}
\usepackage{multirow}
\usepackage{amsmath}
\usepackage{lipsum}
\usepackage{amssymb}
\usepackage{color}
\usepackage{dcolumn}
\usepackage{bm}
\usepackage{natbib}
\usepackage{tabu}
\usepackage{times,graphics}
\usepackage{makecell}
\usepackage{diagbox}

\RequirePackage{mathptmx}      
\RequirePackage{flushend}

\newcommand{\physrep}{{Physics~Reports}}

\def\physrep{Phys. Rep.}

\graphicspath{{./fig/}}

\begin{document}

\title{Reconstruction of dark energy density by non-parametric approaches}

\author{Ahmad Mehrabi }
\affiliation{Department of Physics, Bu-Ali Sina University, Hamedan
	65178, 016016, Iran}

\author{Maryam Vazirnia }
\affiliation{Department of Physics, Bu-Ali Sina University, Hamedan
	65178, 016016, Iran}

\begin{abstract}
The evolution of dark energy density is a crucial quantity in understanding the nature of dark energy. Often, the quantity is described by the so called equation of state, that is the ratio of dark energy pressure to its density. In this scenario, the dark energy density is always positive throughout the cosmic history and a negative value is not allowed. Assuming a homogeneous and isotropic universe, we reconstruct the dark energy density directly from observational data and investigate its evolution through cosmic history. We consider the latest SNIa, BAO and cosmic chronometer data and reconstruct the dark energy density in both flat and  non-flat universes up to redshift $z\sim 3$. The results are well in agreement with the $\Lambda$CDM up to redshift $z\sim 1.5$, whereas all data and methods, in our analysis, provide a negative dark energy density at high redshifts. 
\end{abstract}

\maketitle



\section{Introduction}

After outstanding discovery of the cosmic acceleration \cite{Riess_1998,Perlmutter_1999} via supernovae (SN) data, other independent methods and observations such as the baryon acoustic oscillation (BAO) \cite{Eisenstein_2005,Alam_2017,Gil_Mar_n_2018,deMattia:2020fkb}, the cosmic microwave background (CMB) \cite{Readhead_2004,Tegmark_2004,Spergel_2007,2020}, the large scale structures (LST) \cite{Blake_2011,Hawkins:2002sg} and the weak lensing \cite{Amendola_2008}, all confirmed the discovery. It is generally believed that the source of cosmic acceleration is an unknown energy, so called dark energy (DE). Since then, understanding the nature of DE has been a big challenge for cosmologists.
A simple model, that is consist of around 70$\%$ cosmological constant $\Lambda$ and around 25$\%$ cold dark matter (CDM), provides a very successful model to describe almost all cosmological observations \cite{Peebles_2003,2020}. However, the model suffers from severe theoretical problems which have not been resolved yet \cite{2003PhR...380..235P,perivolaropoulos2008puzzles,padilla2015lectures}.
Furthermore, a very wide range of models including Quintessence and K-essence\cite{COPELAND_2006,2a2da62fc2974cf887c7db2e11d4749f,Amendola:2010}, different kind of holography \cite{Wang_2017,CRUZ2020115017}, dynamical vacuum energy \cite{RevModPhys.61.1,Padmanabhan_2003,padilla2015lectures} as well as modification of gravity \cite{Schmidt_1990,PhysRevD.70.043528} have been proposed to describe the cosmic acceleration.   

Generally speaking, all theoretical models have some free parameters which should be constrained by an observational data. So having a model and some observational data, it is an easy task to perform a statistical parameter inference to obtain the free parameters as well as their uncertainties \cite{Mehrabi:2015kta,Rezaei:2017yyj,Mehrabi:2018dru,Mehrabi:2018oke,Rezazadeh:2020zrd,Rezaei:2020mrj}.

On the other hand, it is possible to study a data set without assuming any parametric form. In such a procedure, the data directly has been used to reconstruct a large number of consistent curves and the method is very useful specifically in investigation of a weird component like DE. the Gaussian process (GP) is one of most well-known methods, utilizing for investigation of cosmological data in \cite{Liao_2019,G_mez_Valent_2018,Pinho_2018,Mehrabi_2020,Dhawan:2021mel,Escamilla-Rivera:2021rbe,Bernardo:2021mfs,Briffa:2020qli,Mehrabi:2021feg}. Moreover, the smoothing method (SM) might be used to reconstruct a curve consistent with a data set. The approach starts from an initial guess and utilizes a smoothing kernel to generate another curve closer to the data points \cite{Shafieloo_2006,Shafieloo_2010,mehrabi2021nonparametric}. The SM method has been used to reconstruct the cosmic expansion history as well as the equation of state (EoS) of the dark energy in \cite{Linder_2003,Kopp_2018,Usmani_2008}.
  
There are generally two avenues to investigation DE properties, through its density or its equation of state (EoS). The EoS is defined as the ratio of the DE pressure to its density $w_{DE}=\frac{p_{DE}}{\rho_{DE}}$. It is worth noting that for a conserved DE fluid with a positive density $\rho_{DE}>0$, the EoS fully determines its evolution. This is the main reason why in lots of works, people consider EoS  instead of density to study the evolution of the DE \cite{Usmani_2008,LAZKOZ2010198}. However, the effective EoS can become singular in some models \cite{khurshudyan2013dark,SOLA2005147} and so the DE density changes its sign and could be negative. It is worth noting that a negative DE density could not be described by a EoS, so investigation of DE in terms of EoS might be incomplete. 
 In this work, we assume an isotropic and homogeneous universe, consisting CDM and a DE component, to reconstruct the DE density directly from the observation. The analysis have been done for both a flat and a non-flat universe. We consider two well-known and basically different cosmological data to reconstruct the DE density up to redshift $\sim 3$. A similar analysis using different methods and data sets has been done in \cite{Wang:2018fng}.

The structure of this paper is as follows:  In section (\ref{sec:density}), we discuss about basic assumptions and present the DE density parameter in both a flat and a non-flat universe. After that in section (\ref{sect:mod-ind}), we briefly introduce and describe the GP and SM, two well-known non-parametric methods, utilizing in this work. The main results of our analysis have been presented in (\ref{sect:result}). We show the reconstructed DE density parameter and compare the results with the concordance $\Lambda$CDM in this section. Finally, we conclude and discuss the main points of our finding in section (\ref{conclude}).

\section{Basic assumptions and dark energy density}\label{sec:density}
In order to reconstruct the dark energy density directly from observational data, some assumptions must be made. The geometry of the universe is a crucial assumption in this regard. Assuming a homogeneous and isotropic universe, the Hubble parameter as a function of redshift is given by:
\begin{eqnarray}\label{eq:hub}
H^2 &=& H_0^2 [ \Omega_{r}(1+z)^4+\Omega_m(1+z)^3+\Omega_k(1+z)^2 \\ \nonumber
&+&\Omega_{\rm DE}X(z) ]
\end{eqnarray}
where $H_0$ is the current expansion rate and $\Omega_{r},\Omega_{m},\Omega_{k}  $, $\Omega_{\rm DE}$ are the density parameters of the radiation, matter, curvature and DE respectively.  The function $X(z)$ describes the evolution of DE and is assumed to be a general function. Note that there is a constrain $\Omega_{k}=1-\Omega_{r}-\Omega_{m}-\Omega_{\rm DE}$ between these coefficients due to the above equation at present time. For the $\Lambda$CDM model, the $X(z)$ is equal to unity throughout of cosmic history and does not change by redshift. In this work, we are going to reconstruct the $X(z)$ function directly from the observational data. apart from some normalization terms, we call this quantity as the DE density throughout of this paper. In addition to the above assumptions, we ignore the radiation component because it is negligible at those redshifts we are going to reconstruct the $X(z)$.  

We use two basically different cosmological data to reconstruct the $X(z)$, and compare the results with the $\Lambda$CDM. The SNIa data provides the distance module as a function of redshift which can be easily converted to the luminosity distance $(d_l(z))$. Given the evolution of luminosity distance, it is straightforward to obtain the Hubble function from following relations \cite{Amendola:2010}.

\begin{eqnarray}\label{eq:hub-from-lom}
&\rm{if}& \Omega_{k}!=0\\ \nonumber
&H (z)& = {{(1+z)[c^2(1+z)^2+\Omega_k H_0^2 d_l^2(z)]^{0.5}} \over {[(1+z) d_l'(z)- d_l(z)]}} \\ \nonumber
&\rm{if}& \Omega_{k}=0\\ \nonumber
&H(z)& = {c\over{{d\over{dz}}{d_l(z)\over{1+z}}}},
\end{eqnarray} 
where c is the speed of light and $'$ denote derivative with respect to the redshift. In this case, the curvature of the universe affects the reconstruction of the Hubble parameter through above equations.

The second data set is the direct measurement of the Hubble parameters. For this data set, we are able to reconstruct the Hubble function directly from the observation and there is no difference between a flat or a non-flat universe. The details of both data sets will be given in (\ref{sect:result}).

For both data sets, we first reconstruct the Hubble parameter as a function of redshift, then the $X(z)$ is given by      
\begin{equation}\label{eq:xz}
X(z) ={ {1 \over{1-\Omega_m -\Omega_k}}{[{{{H^2(z)} \over {H_0^2}}-\Omega_m(1+z)^3-\Omega_k(1+z)^2}] } }.
\end{equation} 
From each reconstructed $H(z)$, it is easy to read the value of $H_0$, hence the equation only depends on $\Omega_m$ and $\Omega_k$. According to the above equation, we are able to obtain the $X(z)$ for each reconstructed $H(z)$ by assuming a value for $\Omega_{m}$ and $\Omega_{k}$. Since the Hubble parameter is obtained from a non-parametric method, the above mentioned procedure will provide a DE density independent of any model. On the other hand, since various observational data constrain the value of $\Omega_m$ very well, we can fix its value and examine DE density for different values of the curvature density.

\section{Model-independent method}\label{sect:mod-ind}
In this section, we briefly review two well-known model independent methods. These methods have been used in our analysis to reconstruct the Hubble parameter and hence the DE density $X(z)$.

\subsection{Gaussian process}   
Given a data set as $(x_i,y_i,\sigma_i)$, the data can be modeled by a GP. The GP is a sequence of Gaussian random variables (RV) at each $x_i$ and provide a unique approach to study the data in a model-independent manner. Since a sequence of RVs can be described by a multivariate Gaussian distribution, a mean function $\mu(x_i)$ and a covariance matrix $\Sigma(x_i,\hat{x_j})$ are needed to build a GP. In fact, the diagonal (off-diagonal) terms in the covariance matrix give uncertainty at each point (correlation between different points). Given the mean function and the covariance matrix, an unknown function $$f(x)\sim GP(\mu(x),K(x,\hat{x}))$$ is modeled by a GP. In this case, many reconstructed curves could be obtained by sampling from the multivariate Gaussian distribution at any set of arbitrary points $(x^{\star})$. To this aim, we only need the mean and covariance matrix that are given by \cite{10.5555/1162254}
\begin{eqnarray}\label{eq:GP}
\mu^{\star} &=& K(x,x^{\star})[K(x,x^{\star})+C_D]^{-1}Y\\
\Sigma^{\star} &=& K(x^{\star},x^{\star}) - K(x^{\star},x)[K(x,x^{\star})+C_D]^{-1}K(x,x^{\star}),
\end{eqnarray}
where $C_D$ is the covariance between observational points and $Y$ is a column vector consist of $y_i$ values. 

A functional form for the kernel is required to compute the covariance function.  There are various kernel functions with different properties for different tasks and the squared exponential is a one of the well-know kernels. The kernel is given by $$K(x,\tilde{x}) = \sigma_f^2\exp{\frac{-(x-\tilde{x})^2}{2\sigma_l^2}}$$, where $\sigma_f^2$ and $\sigma_l$ are two hyper-parameters
 which must be constrained using a Bayesian inference or set to the maximum likelihood values \cite{10.5555/1162254,Seikel_2012}. In this work, we consider exponential kernel and use the {\it{scikit-learn}} library \cite{JMLR:v12:pedregosa11a}to reconstruct the Hubble parameter (luminosity distance) in the case of the Hubble (SNIa) data sets.

\subsection{Smoothing method}   
The iterative smoothing method is another useful approach to reconstruct an unknown function given a data set. The method starts from an initial arbitrary points and tries to find a smooth curve closer to the data points. It only needs a hyper-parameter $\Delta$, depending on the quality and number of data points (to see more details refer to \cite{L_Huillier_2019}). The method has been widely used to study a cosmological data in a model-independent manner. For example, it has been employed to reconstruct the expansion history in \cite{Shafieloo_2012,10.1093/mnras/sty398,Shafieloo_2018}. 

Given a data set $(x_i,y_i,\sigma_i)$ and its covariance $C_D$, the reconstructed function at arbitrary $x^{\star}$ points is given by a iterative relation, 
\begin{equation}\label{eq:scal-fac}
\hat{f}_{n+1}(x^{\star}) = \hat{f}_{n}(x^{\star})+{{\delta f_n^T .C_{D}^{-1}.W(x^{\star})} \over {1^T.C_{D}^{-1}.W(x^{\star})}}, 
\end{equation} 
where the kernel function $W(x^{\star})$ and $\delta f_n^T$ are given by:
\begin{equation}\label{eq:scal-fac2}
W_i(x^{\star}) = exp{   ({{- \ln^2{{(1+x^{\star}) \over{(1+x_i)}} } \over {2 \Delta^2}})} }, 
\end{equation} 
\begin{equation}\label{eq:scal-fac3}
\delta f_n = f(x_i) - \hat{f}_{n}(x_i), 
\end{equation} 
and the $1^T$ is a unite column vector. In this work, we set $\Delta=0.3$ for the SNIa data and $\Delta=0.4$ for the Hubble data. Note that, we have examined other values to check the robustness of our results for each data set and these values are most efficient for our analysis. Starting from an initial guess, the method generates many reconstructed curves, each has a specific $\chi^2$ value. The $\chi^2$ indicates the normalized distance between the reconstructed values $y^{\star}_i$ at $x_i$ and the observations $y_i$,  $$\chi^2 = \sum_i\frac{(y^{\star}_i-y_i)^2}{\sigma_i^2}$$. Now, the process needs a selection criterion to make a sample of reconstructions. Following \cite{Mehrabi:2021uuy}, we use the chi-squared optimization criterion to select a reconstruction and set the probability to $pr=95\%$ (to see more details on the sampling selection see \cite{Mehrabi:2021uuy}).

\section{Results}\label{sect:result}
In order to investigate the DE density in a model-independent manner, we consider two well-know data sets. The first one is a combination of the Hubble parameter measurement from cosmic chronometers as well as BAO measurements. The data has been collected in \cite{Farooq:2016zwm}. We also add the local $H_0$ measurement from nearby SN (SHOES data point) \cite{Riess:2019cxk} to the data set. For this data set, we perform the following steps
\begin{itemize}
	\item A large number of $H(z)$ reconstructions have been generated using both SM and GP methods
	\item For each curve, we compute the $\chi^2$ and select those curves with probability $95\%$ according to the chi-squared distribution 
	\item The value of $H_0$ for each reconstruction in the sample is read at $z=0$
	\item Setting a value for $\Omega_m$ and $\Omega_k$, we compute the $X(z)$ from Eq. (\ref{eq:xz}) 
	\item Now, we have a sample of $X(z)$, the mean (standard deviation) at each redshift is calculated as the central value (uncertainty) at that point 
\end{itemize}

\begin{figure}
	\centering
	\includegraphics[width=9 cm]{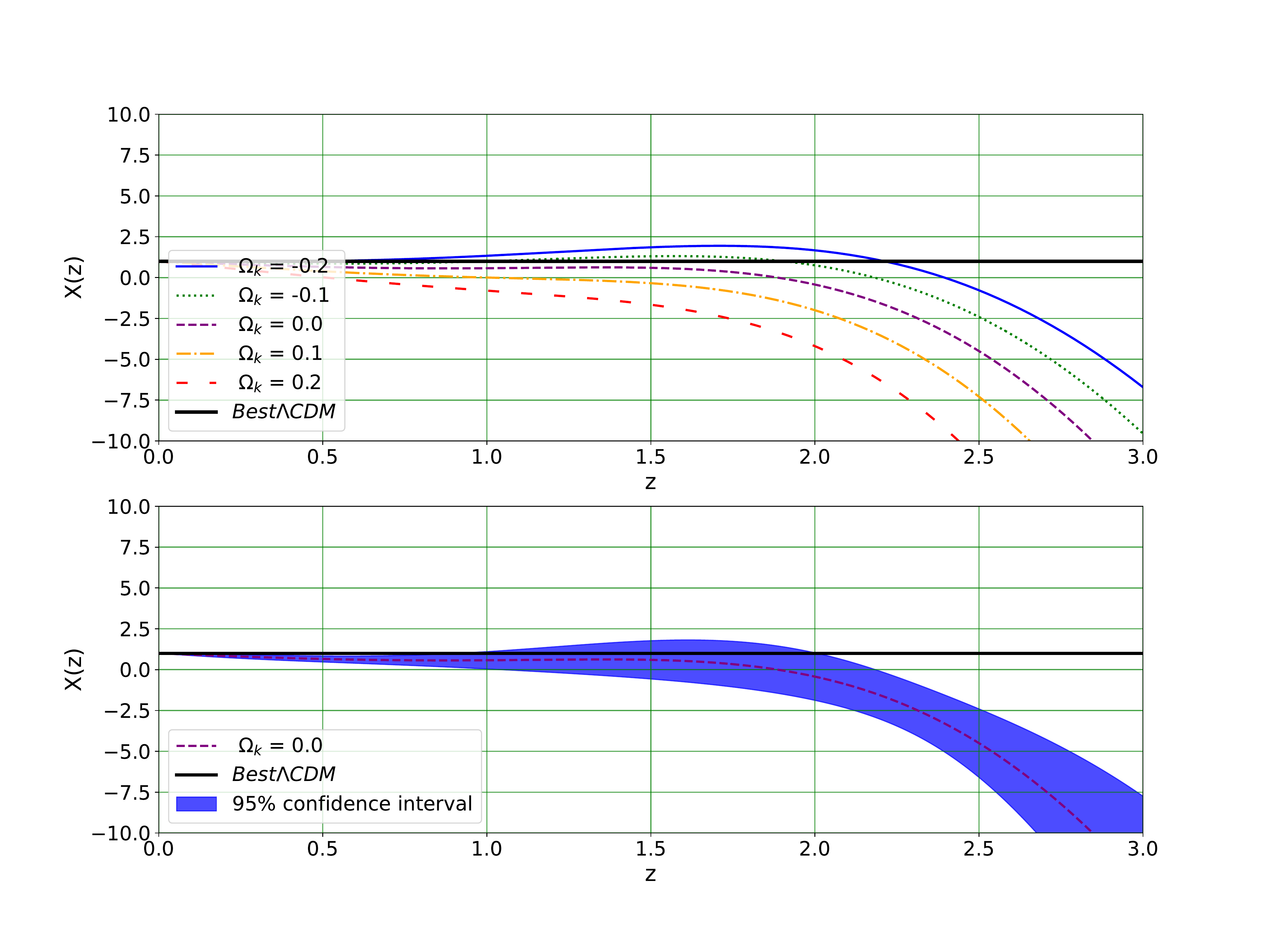}	
	\caption{The results of applying the Hubble parameter data to the GP. The upper panel: the central value DE density as a function of redshift for different values of curvature density. The lower panel: the central value as well as its $95\%$ confidence interval for a flat universe.}
	\label{fig:H_GP}
\end{figure}

\begin{figure}
	\centering
	\includegraphics[width=9 cm]{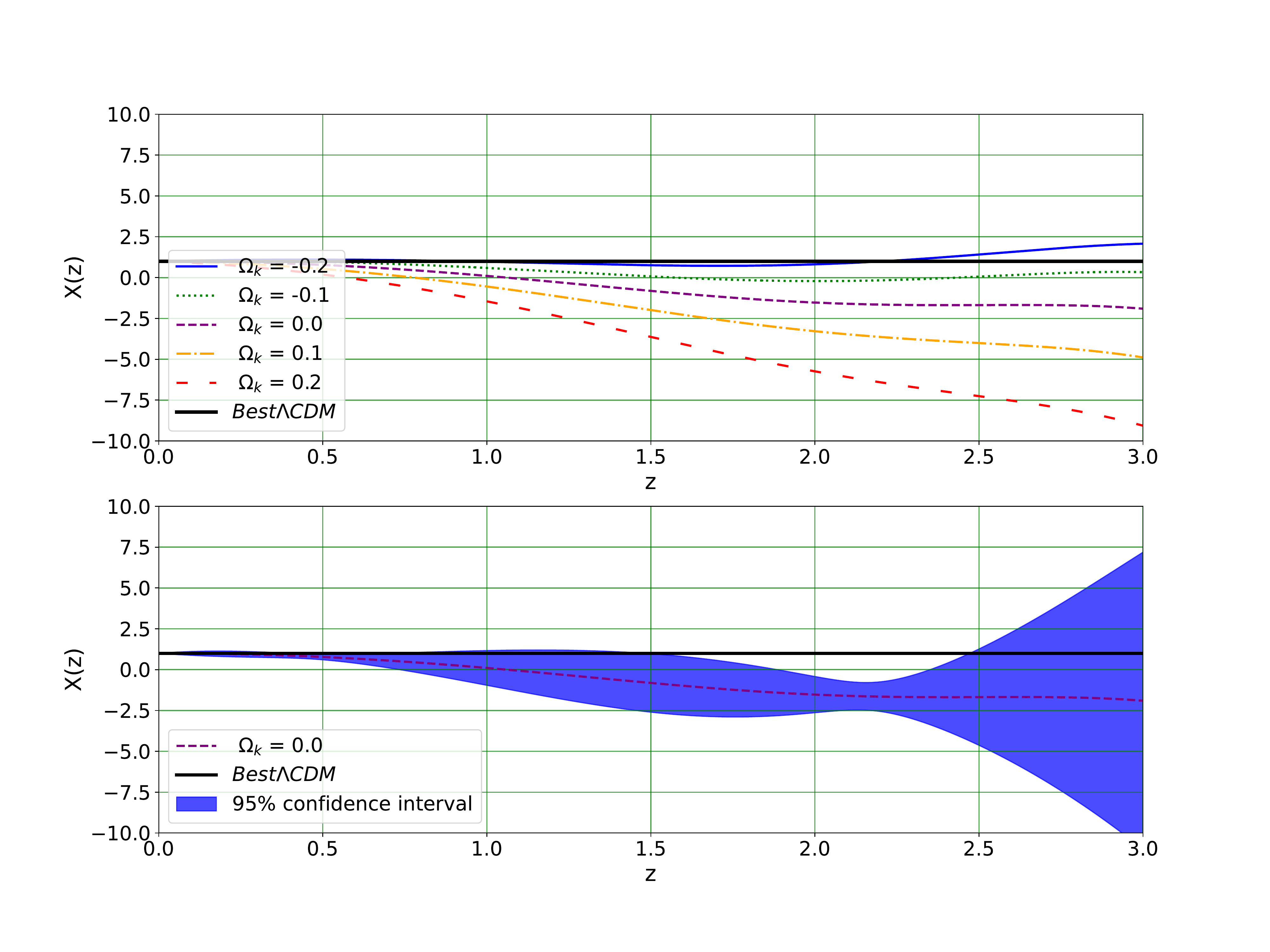}	
	\caption{The results of applying the Hubble parameter data to the SM. The upper panel: the central value of effective density as a function of redshift for different values of curvature density. The lower panel: the central value as well as its $95\%$ confidence interval for a flat universe. }
	\label{fig:H_SM}
\end{figure}

Since a variety of observations indicate $\Omega_m\sim 0.3$, we set the value throughout our analysis. The value has been obtained independently from CMB measurement in \cite{2020}, from BAO in \cite{deMattia:2020fkb} and from SNIa Pantheon sample in \cite{Scolnic_2018}. The results for GP and SM methods have been presented in Figs. (\ref{fig:H_GP}) and (\ref{fig:H_SM}) respectively. 

The upper panel in Fig. (\ref{fig:H_GP}) shows the central values for different curvature densities and the lower panel shows only the flat case along with its $95\%$ confidence interval. Since confidence intervals have a large overlap, we only show one case to provide a better illustration. The black solid line indicates the value of $\Lambda$CDM and results for different curvatures have been presented by different colors and line styles. All the DE densities are consistent with the $\Lambda$CDM up to redshift $\sim 1$ but the deviation increases at high redshifts. It is very interesting that, we see a negative value for the DE density at higher redshifts $z\gtrsim 2$ for all the curvature densities. Moreover, our results indicate that by increasing the curvature density, the DE density decreases and a positive curvature density provides a more negative DE density. For a flat universe $\Omega_k=0$, the DE density is consistent with the $\Lambda$CDM up to redshift $z\sim 2$ at $95\%$ confidence interval but present a negative value for $z>2$. In this case, we obtain X=-0.48( $1.1 \sigma$ deviation from the $\Lambda$CDM) at $z=2$ and X=-13.2( $2.5\sigma$ deviation from the $\Lambda$CDM) at $z=3$.  
 It is worth noting that, our results are in excellent agreement with those presented in \cite{Wang:2018fng}. Authors of \cite{Wang:2018fng} have reconstructed the DE density by a completely different approach and obtained similar results.

The results of SM method, on the other hand, have been presented in Fig. (\ref{fig:H_SM}). Generally speaking, the results are similar to the GP case but the SM provides a relatively larger uncertainty at high redshifts. In addition, the central values do not decrease as sharply as the GP at high redshifts. In this case, for a null or positive curvature, we have a negative DE density at high redshifts but the results for a negative curvature are consistent with $\Lambda$CDM. This is mainly due to the fact that the SM provides a larger uncertainty compare to the GP.

The second data set, we use in our analysis is the SNIa data from the Pantheon sample \cite{Scolnic_2018}. The data consists of 1048 spectroscopically confirmed SN in the redshift range $(0.01<z<2.26)$. Notice that, it is not possible to obtain the Hubble parameter from the SN data unless the value of absolute magnitude is set. For example, a $M_B=-19.4 $ ($M_B=-19.24$) will result in $H_0=68.67$ ($H_0=74.23$). In the current analysis, we set an intermediate value for the absolute magnitude ($M_B=-19.3$).  For the SN data, we perform steps as follows,
\begin{itemize}
	\item The distance module has been converted to the luminosity distance via $D_L = 10^{\frac{\mu-25}{5}}$
	\item Using GP and SM methods, a large number of reconstructed luminosity distance are obtained 
	\item Among all reconstructions, those with $95\%$ probability, according to the chi-squared PDF, are selected to make a sample
	\item Given a value of $\Omega_{k}$, the Hubble parameter for each reconstruction has been obtained using Eq. (\ref{eq:hub-from-lom})
	\item Finally, setting the value of $\Omega_{m}$, the $X(z)$ for each reconstruction is obtained and the central value and uncertainty are computed in a similar procedure as the Hubble data.  
\end{itemize}

\begin{figure}
	\centering
	\includegraphics[width=9 cm]{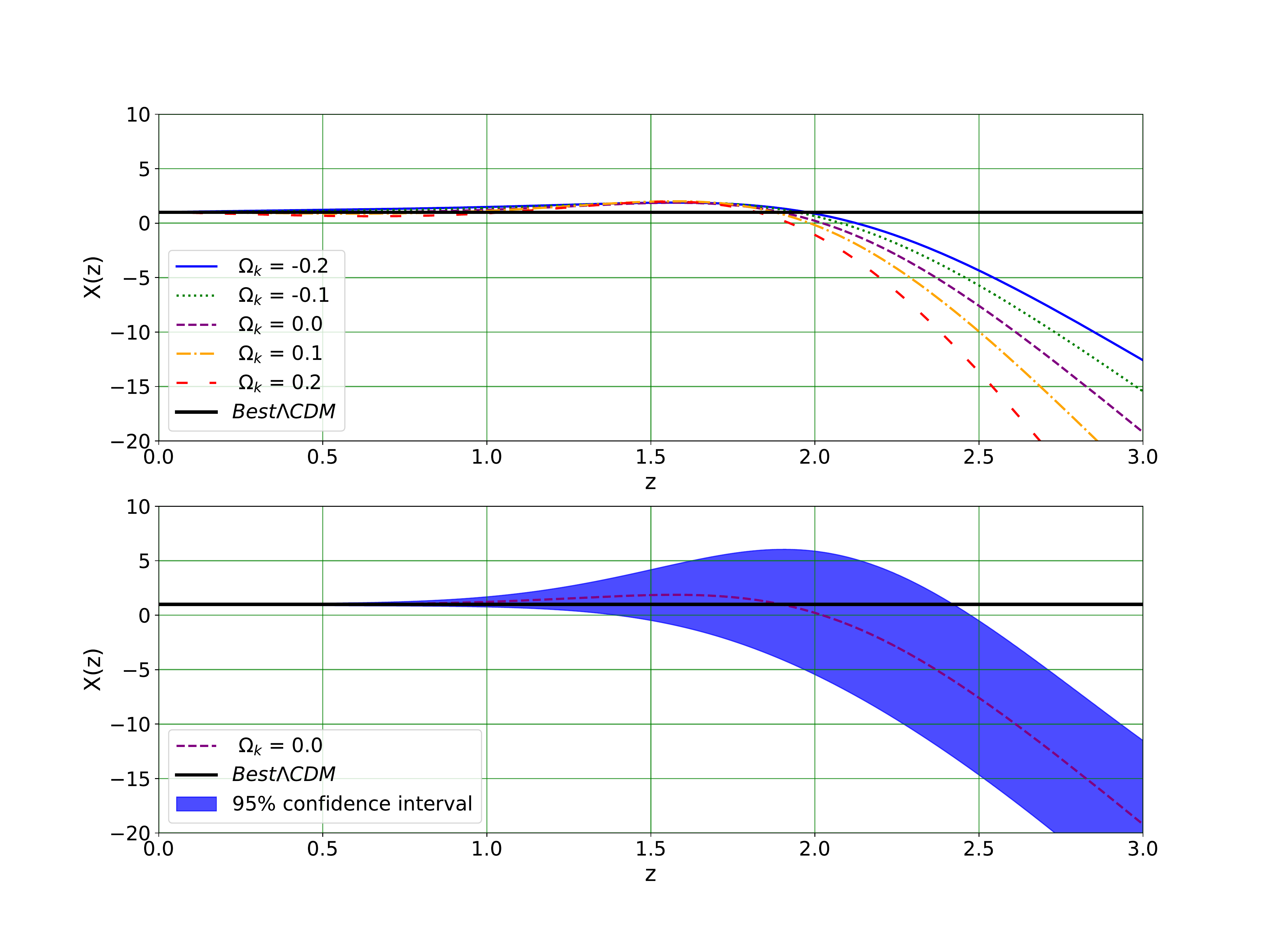}	
	\caption{The results of applying the SN data to the GP. The upper panel: the central value of DE density as a function of redshift for different values of curvature density. The lower panel: the central value as well as its $95\%$ confidence interval for a flat universe.}
	\label{fig:SN_GP}
\end{figure} 

\begin{figure}
	\centering
	\includegraphics[width=9 cm]{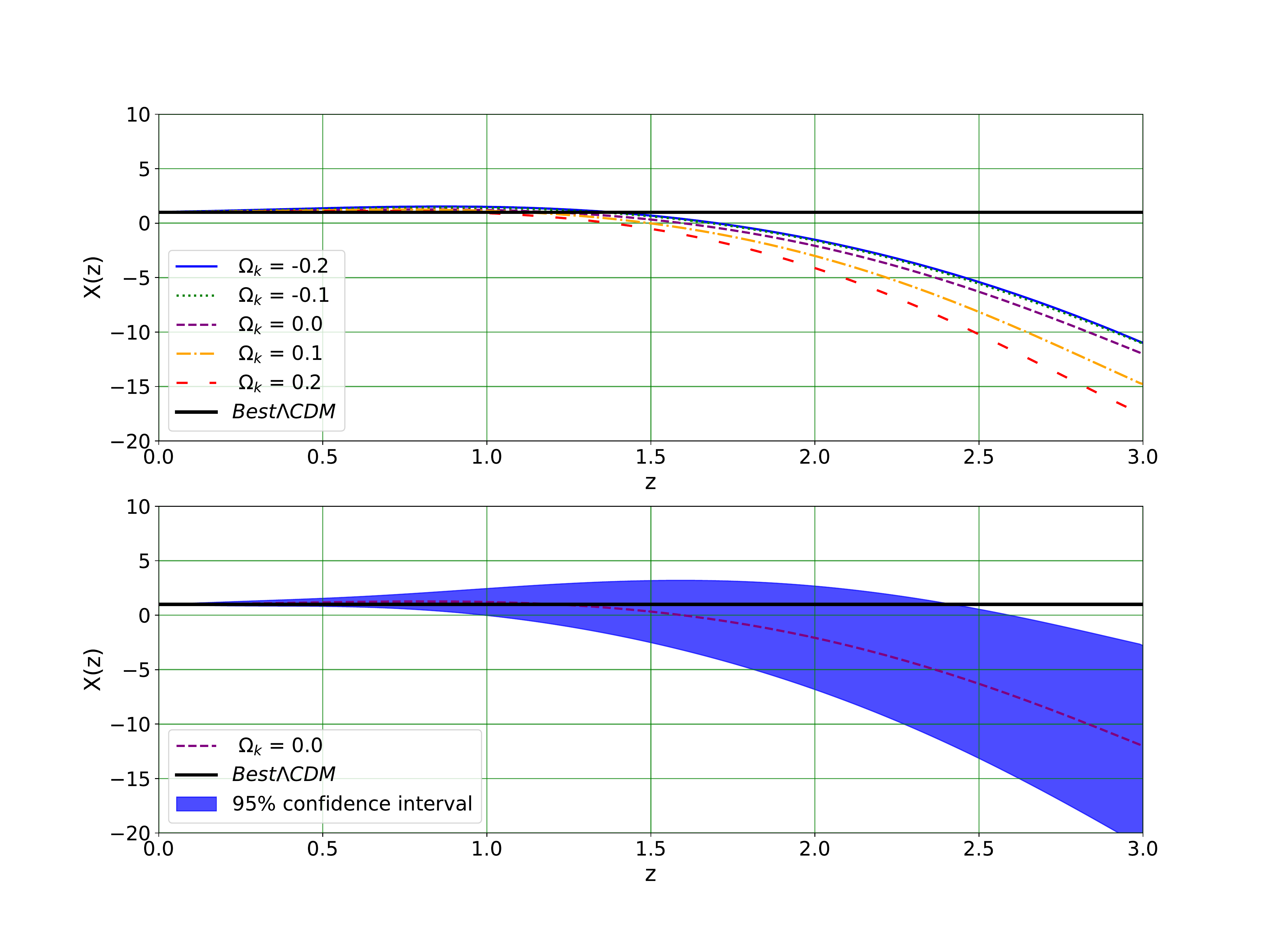}	
	\caption{The results of applying the SN data to the SM. The upper panel: the central value of DE density as a function of redshift for different values of curvature density. The lower panel: the central value as well as its $95\%$ confidence interval for a flat universe. }
	\label{fig:SN_SM}
\end{figure} 

The reconstructed DE densities for different value of curvature density have been presented in Figs. (\ref{fig:SN_GP}) and (\ref{fig:SN_SM}) by considering the GP and SM methods respectively. The results from the GP indicate all reconstructions are consistent with the $\Lambda$CDM up to redshift $z\sim2$ and different curvature density provides almost the same results. In contrary, at high redshifts $z>2$, we see a negative DE density similar to the Hubble data. On the other hand, the results of SM method are almost similar to the GP but the reconstructions start to deceases at $z\sim 1.5$ instead of $z\sim2$. As we mentioned before, the results are very interesting because it is not possible to describe a negative DE density scenario through a EoS. According to our analysis, the negative DE density appears at high redshifts for both data sets. Our findings indicate that describing a DE as a fluid with an arbitrary EoS might not be true, specifically at high redshifts. A similar conclusion has been obtained through different method and data set in \cite{Wang:2018fng}.

\section{Conclusion}\label{conclude}
A non-parametric approach provides a unique tool to study a data set free of any parametric model. These methods have been widely used in study the cosmological data and can provide very useful insights. In this work, we employ two basically different cosmological data, the SN (Pantheon sample) and recent Hubble parameter data, to reconstruct the Hubble parameter as a function of redshift independent of any model. These reconstructions, then, are use to reconstruct the DE density up to redshift $z=3$. To this aim, we use two well-known methods, the GP and SM. For the Hubble data, the Hubble parameter is given directly from the data but for the SN data, first the luminosity distance has been reconstructed and then converted to the Hubble parameter. We consider both flat and non-flat universes to convert the luminosity distance to the corresponding Hubble parameter.

On the theoretical side, a dynamical DE is considered to describe the observational data. Such a model is parameterized in terms of a EoS and its density can't be negative through cosmic history. In contrast to this view point, in some DE models, the DE density can change sign, so the EoS become singular \cite{Deffayet:2009mn,Deffayet:2009wt,Deffayet:2011gz}. Since we know little about nature of DE, investigation of cosmological data free of any parametric model might reveal some new features. Specifically, reconstruction of DE density in a model independent way will be useful to understand the evolution of DE through cosmic history. In this regard, we apply two well-know non-parametric modeling methods and consider two basically different cosmological data to reconstruct the DE density. 

For the Hubble data, both the GP and SM provide a consistent results with the $\Lambda$CDM up to redshift  $z\sim1$ and a negative DE density at high redshifts. It is worth noting that, this trend does not change by varying the curvature density  and both positive and negative curvatures provide a similar pattern. Furthermore, considering the SN data, the GP and SM give a similar results. In this case, the DE density starts to decreases at $z\sim1.5-2$ and shows a negative value at higher redshifts independent of curvature density value. Our finding are very interesting because describing DE in terms of EoS has been widely used in the literature and our results, independent of any model, indicate that this strategy might not work at high redshifts. Similar conclusion, using a different data sets and methods, has been obtained in \cite{Wang:2018fng}.


 \bibliographystyle{apsrev4-1}
  \bibliography{ref}

\end{document}